\documentclass[conference]{IEEEtran}

\makeatletter
\def\ps@headings{%
\def\@oddhead{\mbox{}\scriptsize\rightmark \hfil \thepage}%
\def\@evenhead{\scriptsize\thepage \hfil \leftmark\mbox{}}%
\def\@oddfoot{}%
\def\@evenfoot{}}
\makeatother
\IEEEoverridecommandlockouts
\usepackage[letterpaper, left=0.635in,right=0.635in,top=0.75in,bottom=1.04in]{geometry}

\usepackage{cite}
\usepackage{amsmath,amssymb,amsfonts}
\usepackage{algorithmic}
\usepackage{graphicx}
\usepackage{textcomp}
\usepackage{xcolor}
\usepackage{listings}
\usepackage{lipsum}
\usepackage{multirow}
\usepackage{pifont}
\usepackage{newunicodechar}
\usepackage{enumitem}

\usepackage[hyphens]{url}
\def\BibTeX{{\rm B\kern-.05em{\sc i\kern-.025em b}\kern-.08em
    T\kern-.1667em\lower.7ex\hbox{E}\kern-.125emX}}
\begin{document}


\title{ranDecepter: Real-time Identification and Deterrence of Ransomware Attacks\\
}


\author{\IEEEauthorblockN{Md Sajidul Islam Sajid}
\IEEEauthorblockA{\textit{Computer and Information Sciences} \\
\textit{Towson University}\\
Towson, USA \\
msajid@towson.edu}
\and
\IEEEauthorblockN{Jinpeng Wei}
\IEEEauthorblockA{\textit{Software and Information Systems} \\
\textit{University of North Carolina at Charlotte}\\
Charlotte, USA \\
jwei8@charlotte.edu}
\and
\IEEEauthorblockN{Ehab Al-Shaer}
\IEEEauthorblockA{\textit{Software and Societal Systems Department} \\
\textit{Carnegie Mellon University}\\
Pittsburgh, USA \\
ehab@cmu.edu}
}
\maketitle
\begin{abstract}
Ransomware (RW) presents a significant and widespread threat in the digital landscape, necessitating effective countermeasures. Active cyber deception is a promising strategy to thwart RW and limiting its propagation by misleading it with false information and revealing its true behaviors. Furthermore, RW often acts as a communication conduit between attackers and defenders, allowing deception to return false data to attackers and deplete their resources. This paper introduces \textbf{\textit{ranDecepter}}, a novel approach that combines active cyber deception with real-time analysis to enhance defenses against RW attacks. The ranDecepter identifies RW in real-time and isolates it within a deceptive environment, autonomously identifying critical elements in the RW code to create a loop mechanism. By repeatedly restarting the malware and transmitting counterfeit encryption information and secret keys to the attacker, it forces the attacker to store these fabricated details for each victim, thereby depleting their resources. Our comprehensive evaluation of ranDecepter, conducted using 1,134 real-world malware samples and twelve benign applications, demonstrates a remarkable 100\% accuracy in RW identification, with no false positives and minimal impact on response times. Furthermore, within 24-hours, ranDecepter generates up to 9,223K entries in the attacker's database using 50 agents, showcasing its potential to undermine attacker resources.
\end{abstract}
%
%
%
\section{Introduction}\label{sec:intro}
Ransomware (RW) has emerged as one of the most disruptive and persistent forms of cybercrime. Unlike traditional malware that often prioritizes stealth, RW overtly encrypts user data or locks systems, demanding payment for restoration. These financially motivated attacks indiscriminately target sectors such as healthcare, law enforcement, and critical infrastructure, aiming for rapid disruption and high-impact extortion. The infamous WannaCry outbreak exploited the EternalBlue vulnerability, infecting over 200{,}000 systems worldwide~\cite{WannaCryVictims1}. The simplicity of RW code, combined with its profitability, has fueled its rapid proliferation and increasing sophistication.

Efforts to counter RW span a broad spectrum of defensive strategies. \textbf{Static analysis}, including signature and ML-based techniques~\cite{kharaz2016unveil,baldwin2018leveraging}, can recognize known malware but is easily bypassed through polymorphism, packing, or code obfuscation. \textbf{Dynamic analysis} improves generalization by executing and monitoring malware on real systems~\cite{scaife2016cryptolock,9647816}, but depends on runtime indicators—such as file encryption or deletion—that introduce delays and permit partial damage. ML-enhanced dynamic approaches~\cite{hwang2020two,9647816} train on behavioral traces to model malicious patterns, yet they still rely on post-facto observations and suffer from false positives. Other methods attempt to contain or recover from RW effects. \textbf{Sandboxing} isolates execution environments to observe malware pre-deployment~\cite{alhawi2018leveraging}, but RW can act before redirection or detect sandbox conditions. \textbf{Recovery-based solutions} such as PayBreak~\cite{kolodenker2017paybreak} and caching-based approaches~\cite{elkhail2023seamlessly} attempt to restore encrypted data, but cannot prevent initial file damage and often experience system overhead. 

Deception-based defenses plant \textbf{decoy or honeyfiles}\cite{gomez2018r,moore2016detecting,feng2017poster,mehnaz2018rwguard,wang2018ransomtracer,sheen2022r} throughout the filesystem, triggering alerts when accessed. While lightweight and easy to deploy, these systems are often difficult to manage and prone to false positives. Maintaining separation between honeyfiles and legitimate user activity is nontrivial, especially in dynamic environments, where accidental access or overlap can trigger spurious alerts. Furthermore, these techniques can be bypassed by RW that selectively targets user-specific files. ML-driven decoy generation\cite{ganfure2023rtrap} enhances realism using filesystem metadata, but still relies on passive observation and often requires partial encryption before triggering intervention (see Section~\ref{compareSimilar}).

This asymmetric threat landscape has prompted exploration of defenses beyond conventional monitoring. \textbf{Cyber deception} offers a proactive alternative by disrupting attacker operations through misleading data, manipulated responses, and false artifacts. In the case of RW, deception enables intervention during execution, preventing harm rather than reacting afterward. To be effective, it requires two capabilities: \textbf{accurate runtime identification} of malicious processes and \textbf{timely, precise intervention} before damage occurs. Existing defenses often fail on both fronts, acting too late or misclassifying benign activity. Our system addresses these limitations through \textbf{zero false-positive identification} and \textbf{real-time containment} via API-level deception. Unlike prior defenses that rely on passive monitoring, partial observation of malicious behavior without containment, or post-infection remediation, our system operates directly on production systems—intercepting and containing RW behavior at the API level before any file can be encrypted or deleted. RW behavior consistently follows a recognizable pattern—file enumeration, access, encryption, and deletion. By manipulating these operations at runtime, we invalidate malicious actions without affecting benign applications. This approach is robust to obfuscation and requires no prior knowledge of specific RW variants. It also eliminates reliance on static artifacts or metadata-based heuristics; by leveraging consistent API-level behavior across RW families, it generalizes effectively without exposing sensitive files. Beyond identification and prevention, we demonstrate that deception can be weaponized to increase adversarial cost. Most RW families generate and transmit a unique victim ID and encryption key per infection. By injecting a looping mechanism into the RW binary, we force the malware to repeatedly regenerate and transmit keys, overloading attacker infrastructure and depleting operational resources. This paper makes the following contributions: 

\begin{itemize}[leftmargin=*, noitemsep, topsep=0pt]
  \item We present a proactive, in-host deception framework that performs accurate, zero false-positive identification of RW during early execution, enabling real-time containment and prevention of malicious behavior before any encryption or deletion occurs.

  \item We develop a binary instrumentation technique that locates and modifies key control-flow regions in RW binaries, injecting a looping mechanism that forces continuous key generation and transmission, effectively depleting attacker-side resources.

  \item We evaluate our system using real-world RW samples, achieving 100\% identification accuracy with negligible overhead. With just 50 agents, our system generated up to 9223k database entries per day on the attacker's side.
\end{itemize}

The remainder of this paper is organized as follows: Section~\ref{threatmodel} defines the threat model. Section~\ref{framework} presents the system architecture. Section~\ref{five} describes our evaluation results. Section~\ref{related} discusses related work. Section~\ref{eight} outlines limitations, discussion, and conclusions.

\section{Ransomware Threat Model and System Scope}\label{threatmodel}
\subsubsection{Attack Vectors and Execution} Ransomware (RW) commonly infiltrates systems through phishing emails, malicious downloads, and social engineering. Once it bypasses initial defenses, RW typically encrypts files and exfiltrates associated keys to a remote command-and-control (C2) server in preparation for ransom demands. Despite potentially advanced tactics such as process injection and multi-process execution, the RW still relies on standard operation such as file enumeration, encryption or deletion of data, and network functions for exfiltration. Empirical studies \cite{begovic2023cryptographic,craciun2019trends} indicate that approximately 93\% of RW leverages widely used cryptographic APIs (e.g., Windows CryptoAPI, OpenSSL, Crypto++). This prevalence arises from three primary factors: \textbf{(a)} the proven security of standard algorithms (e.g., AES, RSA, Blowfish, RC4), which resist straightforward cryptanalysis, \textbf{(b)} the high risk of custom cryptographic implementations, historically demonstrated by weaknesses in threats like Apocalypse and GPCode that were rapidly neutralized, and \textbf{(c)} the ability to blend in with benign processes by utilizing common encryption APIs.

\subsubsection{Platform Considerations} Windows is the initial focus of our approach due to its high market share (81.91\%) and the fact that 95\% of RW attacks targeted Windows systems \cite{ransomwareTargets}. Nonetheless, our methodology is not inherently limited to Windows. By adapting the interception and manipulation of equivalent system calls on other platforms—such as POSIX APIs on Linux—and leveraging our platform-independent standard cryptographic library signature detection, our framework can be generalized to detect and disrupt RW across a broad range of operating environments. Our approach targets user-space ransomware; kernel-mode variants that bypass such mechanisms are out of scope and left for future work.

\begin{figure}[htbp]
\centerline{\includegraphics[width=.46\textwidth]{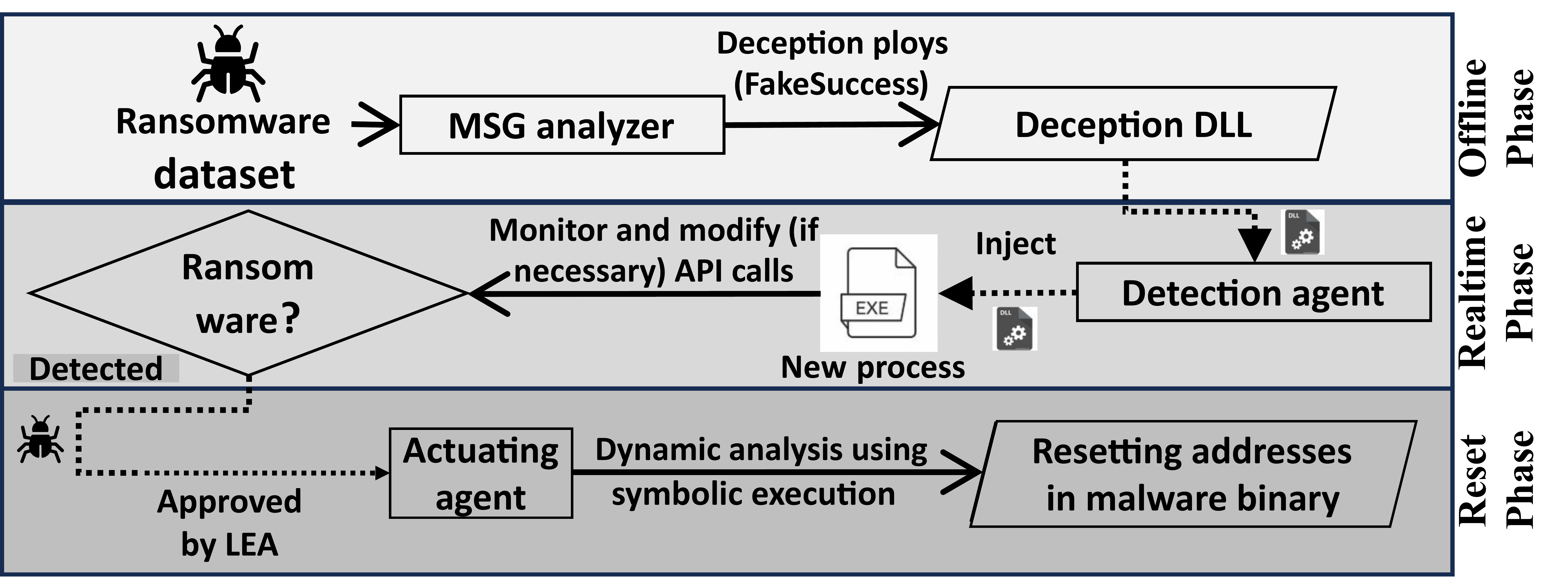}}
\vspace{-3mm}
\caption{Overall system and data flow [LEA: Law Enforcement Agencies]}
\label{fig:systemarchi}
\vspace{-1mm}
\end{figure}

\subsubsection{Assumptions} We assume the RW has already bypassed traditional defenses and is running on the victim machine. Our system is pre-deployed, enabling real-time interception of Windows cryptographic API calls and signature-based detection of standard cryptographic library usage. Finally, we assume attackers must maintain per-infection keys and victim IDs, which renders them susceptible to resource depletion through repeated key generation and storage.

\section{Randecepter: System Overview}
\label{framework}
This section outlines our system architecture (Figure~\ref{fig:systemarchi}), which includes three modules: an offline phase, a real-time ransomware (RW) identification phase, and a binary reset phase. To address legal and ethical concerns, we provide two deployment models. In the first option, the offline and reset phases run on secure infrastructure or servers managed by authorized agencies (e.g., DoD, HSI, CISA), which have the legal authority to undertake countermeasures, while real-time RW identification runs on user systems. In the second, law enforcement manages the full system, deploying real-time modules on honeypots or intentionally exposed systems to ensure legally sanctioned countermeasures.

\subsection{Offline Phase: Ransomware Analysis and Deception Instrumentation for Identification}\label{fourPointOne} This phase builds a behavioral understanding of ransomware (RW) and implements deception strategies for proactive containment and identification. It involves two components: \textbf{(1)} constructing a behavioral knowledge base and \textbf{(2)} deploying deception mechanisms that hook RW's API calls for the real-time phase. RW typically interacts with system APIs for cryptography and file operations. File operations often invoke WinAPI calls like CreateFile and WriteFile, while cryptographic functionality is achieved via three main approaches: (1) dynamic linking of standard libraries (e.g., Windows CryptoAPI, OpenSSL), (2) statically linked libraries (e.g., embedded OpenSSL or Crypto++), or (3) custom cryptographic routines. According to prior surveys~\cite{begovic2023cryptographic,craciun2019trends}, 93\% of RW relies on standard or statically linked cryptographic libraries.

\textbf{Building the Knowledge Base:} We collected and executed 521 RW samples from 30 families (sources: VirusTotal and any.run) in a customized Cuckoo Sandbox to monitor behavior. From these executions, we extracted two foundational artifacts: \textbf{Malicious Subgraphs (MSGs)} and \textbf{Cryptographic Function Signatures (CFS)}. MSGs are data flow graphs where nodes represent API calls and edges encode dependencies. Unlike previous efforts~\cite{sajid2021soda,sajid2023symbSoda} that focused solely on WinAPI, our MSGs also include cryptographic function calls from the standard cryptographic libraries like `CryptEncrypt' or `AES\_encrypt', and file operations such as `CreateFile' and `DeleteFile', capturing complete encryption workflows. In parallel to MSGs, we extract CFS to detect cryptographic activity in statically linked binaries. CFS are compact function-level fingerprints (32 bytes) derived from unique features such as cryptographic constants (e.g., AES S-box), instruction sequences (e.g., aesenc), and control-flow patterns. These signatures enhance offline and runtime identification by enabling efficient matching without reliance on standard library references. For RW employing proprietary cryptographic routines, static analysis is insufficient. In such cases, we apply entropy-based runtime heuristics using tools like \texttt{ent} and \texttt{Volatility}, which detect abnormal entropy patterns in memory and file I/O—reliable indicators of encryption activity. 

Our system generalizes across all cryptographic implementation strategies by supporting: \textbf{(1)} dynamically linked libraries, which are intercepted via API hooking; \textbf{(2)} statically linked libraries, which are identified via CFS; and \textbf{(3)} custom crypto routines, which are identified using entropy heuristics. Additionally, RW must invoke universal file-related APIs (e.g., ReadFile, CreateFile, DeleteFile) to interact with the file system, which our system reliably intercepts.

\begin{table}[!ht]
\centering
\scriptsize
\setlength{\tabcolsep}{4pt}
\renewcommand{\arraystretch}{1.1}
\scalebox{0.9}{
\begin{tabular}{|p{1.6cm}|p{3.3cm}|p{3.9cm}|}
\hline
\textbf{RW Behavior Type} & \textbf{Representative MSGs} & \textbf{Deception Response} \\ \hline

\textbf{File Encryption (Write-to-New-File)} &
CreateFile(O) → CreateFile(D) → Encrypt(CO) → WriteFile(D) → CloseHandle(O, D) → DeleteFile(O) &
Track file handles. Intercept `DeleteFile(O)' and return success without deletion. Preserve original (O). D is flagged for deferred cleanup. \\ \hline

\textbf{File Encryption (Overwrite Original)} &
CreateFile(O) → ReadFile(O) → Encrypt(CO) → SetFilePointer(O) → WriteFile(O) → CloseHandle(O) &
Identify `SetFilePointer(O)' followed by `WriteFile(O)'. Drop write and return success to simulate overwrite while preserving O. \\ \hline
\end{tabular}
}
\vspace{0.1cm}
\caption{Deception ploy planning across encryption and deletion patterns (O: original file, D: destination file, CO: content of original).}
\label{tab:deception_ransomware}
\vspace{-0.4cm}
\end{table}

\textbf{Deception ploys planning:} Our system uses a \textit{FakeSuccess} strategy to deceive RW by simulating successful operations such as encryption, deletion, and overwrite while silently preserving the original files. Table~\ref{tab:deception_ransomware} summarizes two common behavioral patterns observed in RW and outlines the corresponding MSGs and deception responses we deploy.

RW variants that follow the \textbf{Write-to-New-File} pattern first invoke \texttt{CreateFile(D)} to allocate a new encrypted output file, use cryptographic routines such as \texttt{CryptEncrypt}, \texttt{AES\_encrypt}, or \texttt{AES::Encryption} to encrypt content, and write the result using \texttt{WriteFile(D)}. These operations are typically followed by a call to \texttt{DeleteFile(O)} to delete the original file. Our system allows encryption-related API calls to proceed, tracks file handle metadata, and intercepts \texttt{DeleteFile(O)} with a fake success strategy to preserve the original file. These encrypted destination files are flagged for deferred cleanup as described in Section~\ref{detailRP}. 
RW variants using the \textbf{Overwrite Original} strategy skips destination file creation. Instead, it rewinds the file pointer via \texttt{SetFilePointer(O)} and directly overwrites content through \texttt{WriteFile(O)}. Since the goal is to replace the original file, no explicit deletion call follows. Once ranDecepter identifies encryption activity and a write targeting the original file, it drops the write operation and returns success, ensuring the original file remains unchanged. While Table~\ref{tab:deception_ransomware} presents representative MSGs using WinAPI-based RW samples, similar behavioral flows are observed across other libraries. In the case of OpenSSL and Crypto++, the \texttt{CryptEncrypt(CO)} call is replaced by \texttt{AES\_encrypt(CO)} and \texttt{AES::Encryption(CO)}, respectively. Our system handles all such variations uniformly by focusing on the overall behavioral pattern such as whether encryption is followed by a file overwrite or a deletion rather than specific cryptographic APIs.

\begin{figure*}[t]
\centerline{\includegraphics[width=0.86\textwidth]{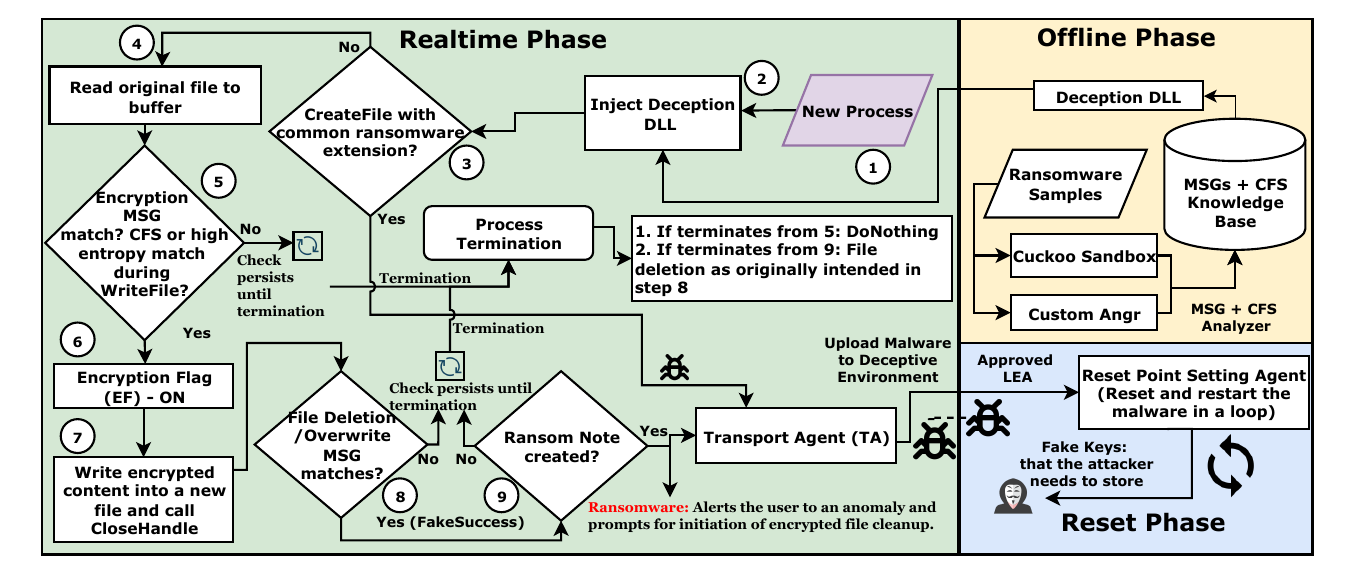}}
\vspace{-12pt}
\caption{Detailed illustration showcasing the various phases and components of our system, depicting their data flow and decision-making processes. [CFR: Cryptographic Function Recognizer; LEA: Law Enforcement Agencies]}
\label{fig:1}
\vspace{-8pt}
\end{figure*}

\textbf{Deception Ploys Implementation (Deception DLL):} To operationalize our deception strategies, we implement a runtime component—the Deception DLL. It hooks critical APIs and injects deceptive responses directly into the RW. As shown in Figure~\ref{fig:1}, this DLL is deployed into live RW processes during execution to intercept encryption, deletion, and overwrite attempts in real time. We leverage the EasyHook library, a flexible open-source engine that supports user- and kernel-level hooking. For our proof-of-concept, we focus on user-mode API hooking, as most RW uses user-level WinAPI calls for encryption and file manipulation. Although only a small subset of RW directly invokes kernel-level APIs for file-related operations, extending our system to the kernel level remains straightforward. This is due to the clear mapping between user-level and kernel-level file operation APIs, combined with EasyHook’s built-in support for kernel-level hooking. Such an extension would further enhance the robustness of our approach, ensuring comprehensive coverage against advanced RW techniques.

\subsection{Realtime Phase: RW Identification using Hooks}\label{detailRP}
Upon process launch, our real-time phase begins by leveraging Windows Defender's WdFilter.sys\cite{WindowsDefender} callback to inject the Deception DLL into new processes (Steps 1 and 2 in Figure~\ref{fig:1}). This ensures immediate interception of system API calls, allowing us to monitor and manipulate ransomware (RW) behavior from the first instruction. Our system offers two runtime policies: \textbf{Whitelisting} and \textbf{Application Restart Control (ARC)} to manage unverified processes during execution. \textit{Whitelisting} determines whether a process should bypass repeated analysis; users may manually add trusted applications or let the system automatically whitelist a process after it runs without triggering RW identification for $n$ consecutive executions. The threshold $n$ is tunable to balance sensitivity with performance, and all whitelist metadata is encrypted and password-protected to prevent tampering. \textit{ARC} controls containment aggressiveness and operates in two modes: \textbf{partial} and \textbf{full} containment. In both modes, attempts to overwrite existing files with encrypted content are intercepted by monitoring \texttt{WriteFile} calls after file pointer resets (e.g., via \texttt{SetFilePointer(0)}). This ensures that original files are preserved regardless of whether the application is ultimately benign or malicious. Since these overwrite operations are not recoverable, benign applications may require a restart to resume normal functionality. In partial containment, encryption to new files is allowed, while deletion operations are blocked and deferred. If the process is later deemed benign, deferred deletions are automatically applied. In full containment, both encryption and deletion are blocked for maximum protection, and benign applications must be restarted if later verified. Users may then choose to whitelist them to avoid future interruptions. To enable precise and adaptive responses, the system employs staged interception and deception strategies that activate dynamically during execution by identifying, containing, and misleading RW without disrupting benign software.

\textbf{Stage 1:} The system first inspects all file creation operations via hooks on the CreateFile API. Specifically, it checks the dwCreationDisposition flag for values such as CREATE\_ALWAYS or CREATE\_NEW, which suggest new file creation. Simultaneously, the lpFileName parameter is parsed to extract the file name and extension, which is then cross-checked against a curated list of known RW file extensions~\cite{ransomext}. If a match is found, the process is immediately terminated and transitioned into the binary reset phase (Step 3, Figure~\ref{fig:1}). While this extension-based approach is highly effective for identifying known RW variants, our system extends beyond simple file extension checks to address behaviors indicative of zero-day RW, which may evade detection by avoiding recognizable extensions. We also hook MoveFile and MoveFileWithProgress, which RW commonly uses to rename encrypted files post-encryption. These APIs are inspected for suspicious renaming behavior, especially involving suspicious file extensions or patterns, enabling early identification of evasive RW activity.

\textbf{Stage 2:} If no suspicious file creation is detected, the system enters Stage 2. Here, the system evaluates whether the process possesses encryption capability. This is determined using three orthogonal techniques introduced in Section~\ref{fourPointOne}: (1) detection of calls to standard cryptographic APIs like CryptEncrypt, AES\_encrypt, or AES::Encryption; (2) runtime memory scanning for Cryptographic Function Signatures (CFS) when WriteFile is called; and (3) entropy-based heuristics using tools like ent and Volatility. When standard crypto API invocations are intercepted, the system turns on the Encryption Flag. For embedded encryption routines, the CFS check is triggered during WriteFile, targeting the memory buffer just before data is flushed to disk—where encryption routines typically complete. This precision avoids full memory scans and ensures lightweight operation. If encryption capability is confirmed and ARC is enabled, the system returns FakeSuccess responses to RW for all sensitive operations from this point forward. If ARC is disabled, encryption is allowed to proceed, with additional safeguards activated in the following stages.

\textbf{Stage 3:} This stage focuses on intercepting how and where encrypted content is written by the process. Two primary scenarios are addressed. In the first, \textbf{new file encryption}, the process creates a new destination file (e.g., using CreateFile(D)) and writes encrypted data via WriteFile(D). In this case, the action is logged and allowed to proceed. The system tracks the destination file(s) and associates it with its handle using CloseHandle(D), marking the file for deferred deletion either upon RW confirmation or process termination. In the second scenario, \textbf{in-place overwrite}, as seen in Case \#2 of Table~\ref{tab:deception_ransomware}, the RW rewinds the file pointer on the original file (SetFilePointer(O)) and attempts to overwrite it via WriteFile(O). If the Encryption Flag is active and ARC is enabled, the write is blocked and a FakeSuccess response is returned to preserve the original content without alerting the RW. If ARC is disabled then the system creates a temporary shadow copy of the original file before allowing the overwrite. If the process is later verified as benign, the copy is deleted; otherwise, the preserved backup protects the user's data from destructive changes.

\textbf{Stage 4:} In the final phase of the new-file encryption model, RW typically invokes \texttt{DeleteFile(O)} to remove the original, unencrypted file after writing the encrypted output to a new file. Upon detecting this call, our system intercepts it and returns ``True", simulating successful deletion. In reality, the original file remains intact, while the newly created encrypted file is marked for deferred cleanup. If the process is later verified as benign, this deletion request is safely re-executed by the system to restore application consistency. This entire phase depends on the Encryption Flag being set—ensuring that these actions are only triggered by processes capable of harm.

\textbf{Stage 5:} In this final stage, the system determines whether the process can be conclusively labeled as RW. A key signal is the presence of a ransom note, which many RW variants generate to demand payment and complete their attack lifecycle. To identify this, we hook the CreateFile API and monitor file creation in sensitive user directories such as Desktop, Documents, and Downloads. Newly created files are scanned for high-risk keywords such as ``ransom," ``encrypt(ed)," ``decrypt(ed)," ``payment," ``bitcoin," ``delete(d)," or ``lose." If such terms are detected, the user is notified to review the file and validate its intent. In addition to textual detection, we monitor visual ransom note cues by capturing background image changes. When changes are detected, the image is processed through Google Cloud Vision API for optical character recognition (OCR) to extract relevant keywords. To further enhance identification accuracy, we incorporate behavior-based techniques from prior work like UNVEIL~\cite{kharaz2016unveil}, and we plan to adopt Latent Semantic Analysis (LSA)~\cite{lemmou2021depth} in future versions to semantically analyze file contents beyond keyword matching.

\begin{table}[ht]
\centering
\scriptsize
\renewcommand{\arraystretch}{1.21}
\setlength{\tabcolsep}{4pt}
\begin{tabular}{|c|p{2.6cm}|p{4.6cm}|}
\hline
\textbf{Stage} & \textbf{Trigger} & \textbf{System Action} \\ \hline
1 & \texttt{CreateFile} with RW extension, or suspicious renaming via \texttt{MoveFile} & If RW extension is detected, terminate process and transition to reset phase. Otherwise, continue monitoring. \\ \hline
2 & Crypto API call, CFS match, or high-entropy detection & Raise Encryption Flag. If ARC enabled, return FakeSuccess to block encryption; else, allow it. \\ \hline
3 & \texttt{WriteFile} targeting new or original file & For new file: allow write and flag for deletion. For overwrite: intercept and FakeSuccess if ARC is on; else copy original. \\ \hline
4 & \texttt{DeleteFile(O)} invoked after encryption & Intercept deletion and return True. If RW confirmed, delete later; else, preserve original. \\ \hline
5 & Ransom note or wallpaper change observed & Confirm RW, initiate cleanup of encrypted files, and finalize classification. \\ \hline
\end{tabular}
\vspace{2mm}
\caption{Staged RW Interception and Deception Logic for Identification}
\vspace{-4mm}
\label{tab:rw-stages}
\end{table}

While RW can deliver ransom messages through multiple channels—including browser pop-ups, lock screens, and email content—a study of 176 RW samples showed that all of them still implemented file-based ransom notes (e.g., \texttt{.txt}, \texttt{.html}) or lock screen modifications~\cite{lemmou2021depth} along with other channels, reinforcing the effectiveness of our current approach. If a ransom note is detected, the process is classified as RW and its binary is uploaded to a secure deceptive environment for analysis and reset-loop activation. If no ransom note is found, but the process exhibits encryption or deletion behavior (e.g., legitimate tools like 7zip), it is not flagged immediately; instead, monitoring continues until process termination to catch delayed ransom note creation. Since deletion and overwrite are already intercepted in Stage 4, this final stage safely resolves the process outcome: if classified as RW, the system removes previously created encrypted files; if benign, all deferred deletion operations are executed as originally intended. This final stage enables confident classification while minimizing false positives and preserving user data throughout execution. The logic described across these five stages is distilled in Table~\ref{tab:rw-stages}, which outlines the key triggers and corresponding system actions. This table serves as a quick reference to our staged interception strategy, helping clarify how ranDecepter achieves identification and response with precision across the RW execution lifecycle.

\subsection{Reset Phase: Depleting Attackers' Resources} This phase demonstrates how ranDecepter’s reset mechanism can support counter-offensive efforts by forcing ransomware (RW) to repeatedly leak its own artifacts such as encryption keys and victim IDs back to the attacker. By automating this loop, ranDecepter increases the adversary’s operational burden, aligning with the goals of national cybersecurity efforts led by organizations like the DoD, CISA, and HSI. While attackers may have scalable storage, our aim extends beyond mere space consumption to significantly increase their operational burden.


\begin{figure}[!h]
\vspace{-10pt}
\centerline{\includegraphics[width=.52\textwidth]{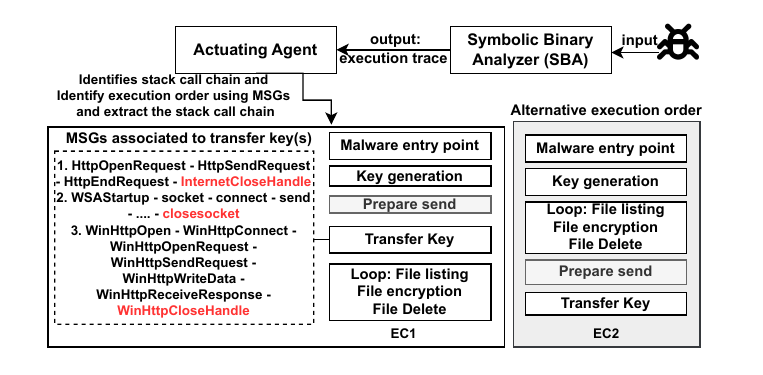}}
\vspace{-15pt}
\caption{Common RW execution chains (EC1 and EC2). On the left side, we present the MSG RW used to perform ``Transfer Key(s)''. Red-colored APIs indicate the end of their respective MSGs.}
\label{fig:opt_msg}
\vspace{-3pt}
\end{figure}

\begin{figure*}[t!]
 \centering
 \vspace{0.7mm}
  \includegraphics[width=.84\textwidth]{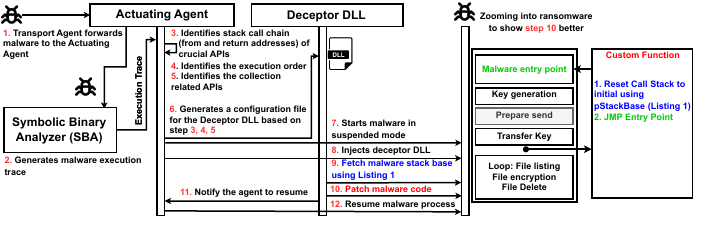}
  \vspace{-10pt}
  \caption{Workflow of binary orchestration (for the execution order EC1).} 
  \label{fig:bin_orc}
  \vspace{-10pt}
\end{figure*}

To implement the looping mechanism, ranDecepter must first locate the malware’s “Transfer Key” stage, where encryption keys and system identifiers are sent to the attacker. This is achieved using the \textbf{Symbolic Binary Analyzer (SBA)}, a custom dynamic analysis module built on gExtractor~\cite{gextractor, sajid2020dodgetron}. SBA employs selective symbolic execution to trace over 390 user-space API calls, log argument values, and capture call stack state, including the caller’s address and API return target. It also extracts the binary’s entry point, critical for redirecting execution into the loop.


Using these traces, the \textbf{Actuating Agent} reconstructs the malware’s execution chain by analyzing the sequence of key operations—such as information gathering, encryption, and network transmission. With help from previously extracted Malicious Subgraphs (MSGs) and Cryptographic Function Signatures (CFS), ranDecepter classifies the RW behavior into two dominant execution orders: \textbf{EC1} and \textbf{EC2} (Figure~\ref{fig:opt_msg}). In EC1, the malware transmits keys before initiating file encryption. In EC2, key transmission occurs after file operations. Recognizing this structure allows precise placement of reset logic. To create a persistent loop, ranDecepter inserts a custom \texttt{JMP} instruction at the return address of the final ``Transfer Key" API (highlighted in red in Figure~\ref{fig:opt_msg}), redirecting execution to the malware’s entry point. To ensure stability, the system restores the original stack state—recorded when the process is initially launched in suspended mode and before re-entry. This includes resetting the stack base and limit to prevent crashes or inconsistent behavior on re-execution. This patching and control redirection is automated using EasyHook. This orchestration is done using \textbf{Deceptor DLL}, injected before malware execution resumes. It has two key responsibilities. First, it contains generic \textit{detour functions} that hook data collection APIs often used to generate victim fingerprints (e.g., MAC addresses, system timestamps, random number generators). These functions are configured to return randomized or misleading data, ensuring each loop iteration produces unique and useless identifiers for the attacker. Second, the DLL manages control flow redirection. Based on a configuration file generated from SBA analysis, it identifies where to insert the reset \texttt{JMP} and how to properly restore the call stack before looping.

The full workflow is illustrated in Figure~\ref{fig:bin_orc}. After SBA completes analysis, the system parses the execution trace logs to locate the end of the key transmission logic. Based on these addresses, it generates a configuration file that specifies randomized data hooks, stack restoration details, and reset jump placement (Steps 3–6). The malware is then started in suspended mode (Step 7), the Deceptor DLL is injected (Step 8), and patching begins (Steps 9–10). Finally, the malware is resumed with a modified control flow (Steps 11–12), entering a loop where it continues to ``infect" the same system and sends fake identifiers repeatedly. In EC1, the loop excludes the encryption and deletion stages, as they occur after the ``Transfer Key" step. However, in EC2, the RW completes file encryption before exfiltration. To handle this, ranDecepter inserts a secondary bypass. It identifies a repeating sequence—starting at `FindFirstFile` and ending at `DeleteFile`—and inserts a \texttt{JMP} from the entry of `FindFirstFile` to the return of `DeleteFile`, effectively skipping the encryption-deletion loop.

\begin{table}[ht]
\centering
\scriptsize
\renewcommand{\arraystretch}{1.05}
\setlength{\tabcolsep}{3.5pt} 
\scalebox{0.96}{
\begin{tabular}{|l r|l r|l r|l r|}
\hline
\textbf{Family} & \textbf{\#} & \textbf{Family} & \textbf{\#} & \textbf{Family} & \textbf{\#} & \textbf{Family} & \textbf{\#} \\
\hline
Bl00dy & 8 & Conti & 19 & PrincessLocker & 13 & Diavol & 14 \\
RUST & 16 & AKIRA & 34 & Anatova & 6 & Lorenz & 3 \\
QazLocker & 8 & NotPetya & 38 & NoEscape & 11 & Jaff & 9 \\
Maze & 108 & RAGNAR & 46 & MedusaLocker & 52 & CryptoMix & 6 \\
AvosLocker & 4 & BadRabbit & 6 & JSWORM & 4 & Mount Locke & 3 \\
CryptoWall & 93 & WannaCry & 121 & dharma & 98 & cerber & 107 \\
lockbit & 97 & cyborg & 38 & saturn & 34 & gandcrab & 110 \\
teslacrypt & 25 & xorist & 3 & & & \textbf{Total} & \textbf{1134} \\
\hline
\end{tabular}
}
\vspace{2mm}
\caption{Representative RW Samples in Our Dataset}
\vspace{-6mm}
\label{tab:dataset}
\end{table}

\section{Evaluations}\label{five}

To evaluate robustness and generalizability, we curated 1,134 ransomware (RW) samples from 30 families collected via VirusTotal and Any.Run, limited to the past five years (details in Table~\ref{tab:dataset}). This dataset, distinct from the one used in Section~\ref{fourPointOne}, was used to assess both real-time identification and orchestration. All samples were executed in a controlled testbed, where API-level interception consistently achieved zero false positives and minimal file loss. To validate resource depletion, we ran four open-source RW samples with accessible C\&C code~\cite{randecepterRepo}, confirming inflated victim ID and key entries in attacker backends.

\subsection{Evaluation of Accuracy and Effectiveness against RW}\label{malEval}

We evaluated ranDecepter’s detection accuracy and overall effectiveness using the 1,134 RW samples from our dataset. Of these, 204 samples were detected early during \texttt{CreateFile} operations by matching known RW file extensions, enabling immediate termination before any damage. The majority (895 samples) delayed renaming until after encryption, using \texttt{MoveFile} or \texttt{MoveFileWithProgress}. For these, ranDecepter leveraged MSG and CFS knowledge to flag suspicious encryption behavior, intercepting and faking subsequent \texttt{DeleteFile} and \texttt{WriteFile} operations via the FakeSuccess strategy to preserve original files. Ransom note creation confirmed RW behavior, triggering alerts with zero data loss. Among the remaining 35 samples, 19 opened network ports but remained inactive due to missing C\&C triggers, while 16 employed custom, embedded crypto routines. These were eventually flagged by entropy-based heuristics, though with an average loss of seven files. Overall, ranDecepter achieved 100\% identification accuracy, effectively detecting RW using static, dynamic, and custom cryptographic techniques. Results are summarized in Table~\ref{tab:stage_vs_loss}.

\begin{table}[!ht]
\centering
\footnotesize
\scalebox{0.93}{
\begin{tabular}{|p{1.0cm}|p{6cm}|p{0.8cm}|}
\hline
\textbf{\# of RW Samples} & \textbf{Identification Stage} & \textbf{Files Lost} \\
\hline
204 & CreateFile with known RW extension & 0 \\
895 & Encryption, File deletion or overwrite, Ransom note & 0 \\
16  & High entropy & Avg: 7 \\
19  & Never reached encryption & -- \\
\hline
\end{tabular}
}
\vspace{2mm}
\caption{Detection Stages vs. File Loss}
\label{tab:stage_vs_loss}
\vspace{-7mm}
\end{table}

\subsection{Evaluation against Benign Applications}\label{benignEval}
To evaluate ranDecepter’s false positive rate and response latency, we tested it against 45 widely used benign applications, including those that exhibit behaviors similar to RW, such as encryption and file deletion. The set included popular programs like Mozilla Firefox, MS Word, 7Zip, WinSCP, VeraCrypt, and MySQL Workbench (see full list in~\cite{randecepterRepo}). Each application was executed using representative workloads; for example, compressing a 125 MB folder with 7Zip.

During evaluation, we executed the benign applications and monitored for false positives, with particular attention to behaviors involving encryption and file deletion or renaming. As expected, no application was misclassified as RW during the \texttt{CreateFile} stage. While several applications performed encryption or file operations, they were not flagged as malicious, since they did not create files with known RW extensions or generate ransom notes. To assess performance impact, we measured each application’s response time with and without our system’s API hooking. All measurements were conducted using WinAppDriver to ensure consistency and automation. As summarized in Table~\ref{tab:app_overhead}, the maximum observed overhead was 5.73\% for 7Zip, followed by 5.44\% for Mozilla Firefox, 4.61\% for WinSCP, 3.79\% for Opera, and 3.69\% for MySQL Workbench. Due to space constraints, the top ten applications with the highest overheads are shown; the complete evaluation results for all 45 applications are available in~\cite{randecepterRepo}. The remaining applications experienced overheads ranging from 0.40\% to 2.99\%, demonstrating that ranDecepter imposes minimal performance degradation in practice.

\begin{table}[!ht]
\centering
\scalebox{0.92}{
\begin{tabular}{|l|c|l|c|}
\hline
\textbf{Application} & \textbf{Overhead (\%)} & \textbf{Application} & \textbf{Overhead (\%)} \\
\hline
7zip              & 5.73 & WinRAR           & 3.41 \\
Firefox           & 5.44 & Picasa           & 3.19 \\
WinSCP            & 4.61 & Calibre          & 3.14 \\
Opera             & 3.79 & Foxit PDF Reader & 3.08 \\
MySQL Workbench   & 3.69 & Photoshop        & 3.06 \\
\hline
\end{tabular}
}
\vspace{1mm}
\caption{Runtime Overhead of Common Applications}
\label{tab:app_overhead}
\vspace{-3mm}
\end{table}

\subsection{Comparison with Existing Similar Work}
\label{compareSimilar}
We compare \textit{ranDecepter} against leading deception-based RW defenses—RWGuard~\cite{mehnaz2018rwguard}, R-Sentry~\cite{sheen2022r}, and R-Trap~\cite{ganfure2023rtrap}—which similarly leverage decoys and behavioral monitoring for identification. We exclude static, signature-based, and sandboxing techniques due to their inherent limitations as discussed in Sections~\ref{sec:intro} and~\ref{related}. Our approach aligns more closely with dynamic, runtime monitoring systems deployed directly on the victim machine. These prior dynamic approaches often suffer from delayed detection and false positives, typically reacting only after partial file damage. While decoys aim to limit impact, they can be inadvertently accessed by benign processes, triggering false alerts. \textit{ranDecepter} addresses these issues through a decoy-free, ML-free design combining three complementary techniques: (1) runtime extraction of MSGs to capture encryption workflows via API dependencies, (2) static CFS to detect embedded crypto routines, and (3) entropy-based heuristics to detect custom encryption logic. It further employs a \textbf{two-phase validation strategy}, preemptively intercepting destructive calls like \texttt{WriteFile}/\texttt{DeleteFile}, and confirming RW intent via staged indicators such as ransom note creation. This layered design ensures robust, real-time containment without false positives, even under aggressive benign activity. We evaluate all systems in terms of file loss, false positives, and operational cost. 

\begin{table*}[h]
\centering
\scriptsize
\vspace{2.5mm}
\begin{tabular}{|l|c|c|c|c|c|}
\hline
\textbf{System} & \textbf{False Positives} & \textbf{Validation} & \textbf{ML/Heuristic Cost} & \textbf{Decoy Mgmt} & \textbf{File Loss} \\
\hline
ranDecepter & None & Ransom note check & None & Not needed & 0 files (98.5\%), $\leq$7 files (1.5\%) \\
R-Sentry    & Possible & None & None & Required & Up to 10 files \\
R-Trap      & Possible & None & ML + regeneration & Required & Avg. 18 files \\
R-Locker    & Possible & None & None & Required & Partial, not reported \\
RWGuard     & Possible & CryptoAPI + behavior & Random Forest ML & Required & Avg. 288 files \\
\hline
\end{tabular}
\vspace{1mm}
\caption{Comparison of ranDecepter with Existing RW Defenses}
\vspace{-7mm}
\label{tab:comparison}
\end{table*}

\textbf{RWGuard} uses decoy file monitoring, CryptoAPI hooking, and a process-level anomaly detector based on I/O activity. However, its reliance on standard APIs and coarse behavioral metrics results in significant file loss, averaging 288 legitimate files encrypted before detection. In contrast, \textbf{ranDecepter} prevents file loss in 98.5\% of cases and limits it to just 7 files in the remaining 1.5\%, without requiring model training or process profiling. \textbf{R-Sentry} and \textbf{R-Trap} both employ decoy files. R-Sentry places honeyfiles based on assumed directory traversal, while R-Trap uses machine learning to adaptively clone user files. Although more resilient than static decoys, both suffer false positives when benign applications access decoys. R-Sentry allows up to 10 files to be lost before alerting, and R-Trap averages 18. R-Trap also incurs moderate overhead from continuous decoy regeneration and ML-driven selection. In contrast, \textit{ranDecepter} requires no decoy or ML infrastructure, avoids false positives, and achieves near-zero file loss with lightweight operation. \textbf{R-Locker} introduces a FIFO-based trap file in user directories, terminating any process that accesses it. This assumes RW follows breadth-first traversal and encounters the FIFO early, which may not hold for targeted RW. Like other decoy-triggered systems, it lacks confirmation logic, increasing the risk of benign interference. Table~\ref{tab:comparison} summarizes these systems across key deployment metrics: false positives, operational overhead, and file loss.


\subsection{Accuracy and Performance Analysis of Binary Orchestration in the Reset Phase}

To evaluate the efficacy of our binary orchestration strategy, we tested it on four open-source RW samples with accessible server-side components: NekRos, Cryptonite, Jasmin, and a POC Windows crypto RW (denoted as r1–r4)~\cite{randecepterRepo}. These samples enabled end-to-end validation of our reset mechanism. 

Each sample was compiled and analyzed within our Symbolic Binary Analyzer (SBA) to extract encryption-relevant API addresses. The Actuating Agent then generated the corresponding Deceptor DLL configurations and injected them into the malware processes. As the binaries executed, server logs confirmed repeated receipt of unique victim IDs and encryption keys, confirming 100\% accuracy in binary orchestration. For performance assessment, we monitored the attacker’s database over 24 hours to assess resource depletion. Our system generated \textbf{3,081K to 9,223K entries}, consuming \textbf{385 MB to 1.1 GB} of memory. Using VPNs/proxies could further enhance this by masking traffic origins. Results are summarized in Table~\ref{tab:db_depletion}.

\begin{table}[!ht]
\centering
\scalebox{1.0}{
\begin{tabular}{|c|c|c|}
\hline
\textbf{Sample} & \textbf{Extra DB Entries} & \textbf{Extra Space (MB)} \\
\textbf{} & \textbf{(1h / 12h / 24h)} & \textbf{(1h / 12h / 24h)} \\
\hline
r1 & 290K / 3420K / 6910K & 43 / 527 / 1054 \\
r2 & 131K / 1582K / 3081K & 15 / 189 / 385 \\
r3 & 413K / 4829K / 9223K & 52.9 / 603 / 1189 \\
r4 & 195K / 2257K / 4357K & 22.7 / 272 / 544 \\
\hline
\end{tabular}
}
\vspace{1mm}
\caption{Attacker-Side DB Misinformation Depletion}
\label{tab:db_depletion}
\vspace{-6mm}
\end{table}


\section{Related work}\label{related}
Ransomware (RW) defenses fall into two broad categories: \textbf{non-deception-based} and \textbf{deception-based}, each with notable limitations that \textit{ranDecepter} overcomes via real-time API-level containment and adversarial disruption.

\textbf{Non-Deception-Based Methods:}
Static and ML-based detection approaches match malware signatures or learn code-level features~\cite{kharaz2016unveil,baldwin2018leveraging}, but are easily evaded by packing, polymorphism, or obfuscation. Dynamic approaches have been proposed to monitor runtime behaviors such as I/O activity or encryption patterns~\cite{scaife2016cryptolock,continella2016shieldfs}, however these systems require observable damage (e.g., file modification or deletion) to trigger alerts, resulting in detection delays and file loss. ML models applied to dynamic traces~\cite{hwang2020two} improve generalizability but continue to depend on post-encryption artifacts and often incur false positives under benign workloads. Sandboxing techniques~\cite{alhawi2018leveraging} aim to isolate malicious samples before deployment; however, many RW strains now include sandbox evasion capabilities and their reliance on isolated execution renders them impractical in real-time, in-place scenarios. Post-hoc tools like PayBreak~\cite{kolodenker2017paybreak} or caching-based approaches~\cite{elkhail2023seamlessly} attempt recovery rather than prevention, leaving files partially encrypted and adding storage overhead.

\textbf{Deception-Based Methods:}
Deception techniques attempt early detection by populating the filesystem with decoy to lure RW activity~\cite{mehnaz2018rwguard,sheen2022r}. These approaches, while conceptually proactive, face several challenges. First, they require wide deployment of synthetic artifacts, which may overlap with legitimate user workflows, introducing false positives~\cite{gomez2018r,wang2018ransomtracer,ahmed2025spade}. Second, adaptive RW can selectively avoid decoys, targeting specific files.
To address these limitations, recent systems such as R-Trap~\cite{ganfure2023rtrap} incorporate metadata-aware cloning and ML-driven decoy placement to enhance realism. However, these remain reactive, as action is only triggered after partial encryption has occurred. For instance, RWGuard, R-Sentry, and R-Trap collectively permit between 10 and 288 file losses before mitigation~\cite{mehnaz2018rwguard,ganfure2023rtrap,sheen2022r}. Furthermore, all these techniques depend on file system-level indicators, limiting scalability and exposing them to evasion via metadata manipulation.

\textbf{Our Approach:}
Unlike prior efforts, ranDecepter is designed to address these limitations by operating at the API level, intercepting RW behavior before any file access occurs. It generalizes across variants by identifying common API usage patterns (e.g., CreateFile, WriteFile, DeleteFile) and applies deception in real-time to suppress encryption or deletion attempts without affecting benign activity. Unlike decoy-dependent systems, we require no synthetic files, no post-encryption indicators, and no user metadata. This ensures zero false positives and complete containment from the first interaction. Additionally, we introduce a binary orchestration mechanism that instruments live RW binaries with a loop, forcing them to repeatedly regenerate and transmit encryption keys. This not only neutralizes the threat but also actively burdens the attacker's infrastructure—a feature absent in prior work. Our API-level approach is thus more scalable, evasion-resistant, and operationally disruptive than existing RW defenses.

\section{Discussion and Conclusion}\label{eight}

This work presents ranDecepter, a proactive deception-based framework for RW defense that enables real-time identification and containment without relying on static signatures, sandbox isolation, or recovery-based methods. By intercepting and manipulating critical API-level operations at runtime, ranDecepter lets RW reveal its behavior without affecting system files. Our evaluation demonstrates \textbf{100\%} identification accuracy with zero false positives across 1,134 real-world RW samples from 30 families. The API-centric design ensures scalability and low overhead. ranDecepter's architecture is agnostic to specific RW variants or file structures, making it suitable for real-world deployment. A novel contribution of ranDecepter is its reset phase, which leverages deception offensively. By analyzing RW binaries and inserting control-flow redirection loops, it forces malware to repeatedly transmit fake encryption keys and victim IDs. With just 50 agents, the system generated over 9 million entries in an attacker’s backend within 24 hours, significantly increasing adversarial cost. This elevates ranDecepter from prevention to active counter-offense, aligning with CISA, DoD, and HSI missions to disrupt attacker infrastructure.

While our approach demonstrates promising outcomes, certain limitations must be acknowledged. Advanced RW may bypass user-space API hooks by verifying or unhooking IAT/EAT entries. To counter this, we plan to incorporate more resilient techniques such as runtime obfuscation~\cite{islam2025secure}, kernel-level hooks~\cite{javaheri2018detection}, and execute-only memory protections using Extended Page Tables (EPT). These methods make hook detection more difficult but require careful tuning to avoid performance overhead. Although our resource depletion strategy raises the attacker's operational costs, sophisticated adversaries may adapt. Nonetheless, the goal extends beyond storage exhaustion by enabling proactive measures like beaconing to reveal C\&C infrastructure, automated sinkholing, and malware poisoning. These tactics provide both disruption and actionable intelligence for law enforcement. We also acknowledge possible unintended consequences, such as victims losing access if attacker infrastructure is dismantled post-payment. However, our focus is on disrupting active RW campaigns, consistent with botnet takedown strategies. Additionally, ranDecepter currently emphasizes parent-child process monitoring. This intra-process focus may overlook RW coordinating across independent processes. Future work will explore inter-process correlation techniques, such as policy-based process correlation~\cite{bidoki2017pbmmd}, and lineage analysis, to enhance detection of distributed RW behavior.

In conclusion, ranDecepter advances the state of RW defense by combining early identification with active deception. It neutralizes threats before damage occurs and depletes attacker resources over time. This proactive paradigm shifts the defense posture from reactive mitigation to adversary disruption. As RW threats evolve, solutions like ranDecepter will be essential for protecting digital assets, critical infrastructure, and organizational resilience in an increasingly hostile cyber landscape.

\bibliographystyle{splncs04}
\bibliography{IEEEexample}

\end{document}